# Influence of Graphene Wettability on Spreading Patterns of Nanodroplets Impacting at High Speed

*Ygor M. Jaques*[1,2] *and Douglas S. Galvao*[1,2*]

[1]Applied Physics Department, University of Campinas – UNICAMP,13083–859–Campinas, Sao Paulo, Brazil

[2]Center For Computational Engineering & Sciences, University of Campinas, Campinas, Sao Paulo, Brazil

**Abstract**

We report here a fully atomistic molecular dynamics study on the dynamics of impact of water nanodroplets (100 Å of diameter) at high velocities (from 1 up to 15 Å/ps) against graphene targets. Our results show that tuning graphene wettability (through parameter changes) significantly affects the structural and dynamical aspects of the nanodroplets. We identified three ranges of velocities with distinct characteristics, from simple deposition of the droplet to spreading with rebound, and finally droplet fragmentation. We also identify that in an intermediary velocity of 7 Å/ps, the pattern of spreading critically changes, due to formation of voids on droplet structure. These voids affect in a detrimental way the droplet spreading on the less hydrophilic surface, as it takes more time to the droplet recover from the spreading and to return to a semi-spherical configuration. When the velocity is increased to values larger than 11 Å/ps, the droplet fragments, which reveals the maximum possible spreading.

**Introduction**

Graphene is one of the most important nanomaterials and it has been extensively studied[1-4] since its experimental realization[5]. Many of its properties, such as structural, electronic, thermodynamic and mechanical[6-10] ones, are already well understood and have been exploited in many different applications. However, in spite of the large number of theoretical and experimental works on graphene, there are still some aspects lacking better understanding. One example is its wettability[11-14].

Experimental works[15-17] tried to determine whether the degree of wettability of graphene is dependent on the substrate, number of layers[18] and/or contaminants[16,19]. It was shown[20] that graphene has only a low adhesion work with liquids like water, formamide, glycerol, among others. A droplet contact angle of 127° for the interaction graphene-water, showing that graphene is hydrophobic, was obtained experimentally[20]. However, later works challenged this interpretation arguing that the hydrophobicity was in fact due to contamination by hydrocarbons in the air[16,19]. When graphene is synthesized in a high vacuum environment, the contact angle is much smaller[16] (as low as 37°). This same work[16] also reported that the hydrophilicity character could be obtained even for graphene exposed to air, but only to a few hours.

Alongside experiments, computational works also studied interactions of liquids with graphene. The determination of parameters that best describe the interaction between the solid and the liquid[21-23] is one of the areas explored by simulations, usually using nanodroplets. Regarding this matter, one of the most used set of parameters for graphitic-carbon and water is the work from reference[24], which gives a contact angle of about 86° for graphene. This early study was carried out assuming graphene had a more hydrophobic character (the discussion about hydrocarbon contaminants started only a few years later). However, a more recent work[22] took into account these contaminations and parameterized the graphene-water interaction with a set of parameters to best fit a hydrophilic character (contact angle of about 40°) for graphene. Besides parameterizations, works involving wettability of graphene usually focus on effects of substrate[18], diffusion of liquid on surface[25] and morphology effects[26,27].

Droplet impacts on surfaces is an area of nanofluidics that has been studied over a century[28], not only because of its fundamental science but also due to its applications in industry, such as inkjet printing, high-pressure spray cleaning, droplet interactions in engines and plasma spraying[29-33], etc. Some of these applications consider droplets impacting the surface at relatively high velocities[34,35] (in the order of 1 Å/ps). Besides that, droplets at even higher velocities (from 5 to 40 Å/ps) can be shot against substrates by the technique of impact desolvation of electrosprayed microdroplets[36,37], which is used to dissolve proteins contained in those droplets.

Few computational studies addressed the impact of droplets on surfaces at this range of high velocities[38-45]. Understanding how water droplets impact on a single layer of graphene is of utmost importance for the applications of this nanomaterial, as well as, to better understand the hydrophilic/hydrophobic behavior of layered materials. However, a detailed study of how the degree of wettability of graphene influences the nanodroplets dynamics is still lacking for such high velocities[41]. This is one of the objectives of the present work. We carried out fully atomistic molecular dynamics (MD) simulations of nanodroplets impacting on graphene surface at different velocities (from 1 Å/ps up to 15 Å/ps). Our results show that the degree of graphene wettability is of fundamental importance to determine the maximum spreading of the droplets. Besides that, distinct patterns of spreading result when the velocity is in the range of 6 up to 11 Å/ps, when structural voids destabilizes the droplet. Increasing the velocity even more leads to the droplet fragmentation.

**Simulation Details**

The MD simulations were carried out with the LAMMPS code[46]. We considered systems composed of graphene sheets (area of 500 x 500 Å[2]) and water droplets with initial diameters of 100 Å.

Each system was first equilibrated for 1000 ps in a canonical ensemble (NVT), using a Nosé-Hoover thermostat[47,48], at 300K. After that, the droplets were shot against the graphene membranes with velocities ranging from 1 up to 15 Å/ps (Figure 1). This velocity range was chosen because it can be experimentally realized[36] and also because they were used in many theoretical papers[38–45], thus allowing direct experimental and theoretical data comparision.

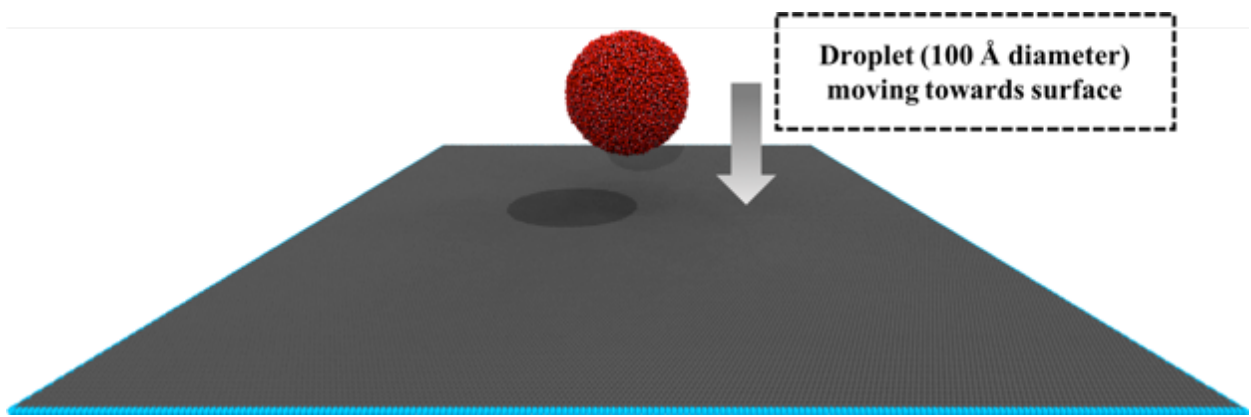

**Figure 1.** Initial configuration of the system consisting of a water nanodroplet shot against a graphene membrane.

After the equilibration process the impact MD simulations were carried out using a micro-canonical ensemble (NVE). As under impact there is a fast change in the kinetic energy (thus, also temperature), the NVE is the natural choice for studying the impact process. A time step of 1 fs was used. The electrostatic interactions were calculated with the Particle-Particle-Particle-Mesh method[49]. The SPC/E model[50] was used to describe the water molecule, and for graphene the AIREBO[51] potential was used. To simulate a droplet impact on the non-rigid graphene surface, the borders of the monolayer (blue in Figure 1) were restrained with a force of 10 kcal/mol in all directions.

We investigated the effect that graphene wettability has on droplet dynamics by using two sets of parameters for the interactions carbon-water. The first set, described here as "more hydrophilic", was obtained from reference[22] and provides a more hydrophilic behavior for graphene (equilibrium contact angle around 40º for nanodroplets). For this model, the interaction between carbon and water is modeled by a Lennard-Jones potential, using parameters: $\varepsilon_{co}$ = 0.0850 kcal/mol, $\sigma_{co}$ = 3.436 Å, $\varepsilon_{CH}$ = 0.0383 kcal/mol and $\sigma_{CH}$ = 2.690 Å. The second set, described here as "less hydrophilic", was obtained from reference[24] and provides a less hydrophilic behavior for graphene (equilibrium contact angle around 86º for nanodroplets). For this set the interaction between carbon and water was modeled by a Lennard-Jones potential as well, using parameters: $\varepsilon_{co}$ = 0.0937 kcal/mol, $\sigma_{co}$ = 3.19 Å, $\varepsilon_{CH}$ = 0.0 kcal/mol and $\sigma_{CH}$ = 0.0 Å.

Surface density maps of the droplets at maximum spread were obtained dividing the simulation box into several bins of size 3 Å and then calculating the density of atoms at each volume. Density maps were also used to calculate the equilibrium contact angle of the droplets after impact, when the droplet potential energy is stabilized. In this case, the cylindrical binning method[21] was used to identify the vapor-liquid interface. Finally, to identify which of the molecules belong to the droplet, excluding the evaporated ones at each frame, we considered a cluster of atoms when the distance between each atom was equal or smaller than 3.3 Å. This approach makes it possible to estimate the temporal evolution of the various droplet properties as its diameter, density and energies.

**Results and Discussion**

*Structural Properties*

For the configurations considered here, the droplets reach graphene surface in a time span of 100 ps, with the spreading depending on the carbon-water interactions and the substantial high kinetic energy that came from the additional downward velocities. As can be seen in the snapshots of Figure 2(a, b), the more hydrophilic parametrization for graphene allows the droplets to spread more at surface in comparison to the less hydrophilic parametrization case at the same impact velocity.

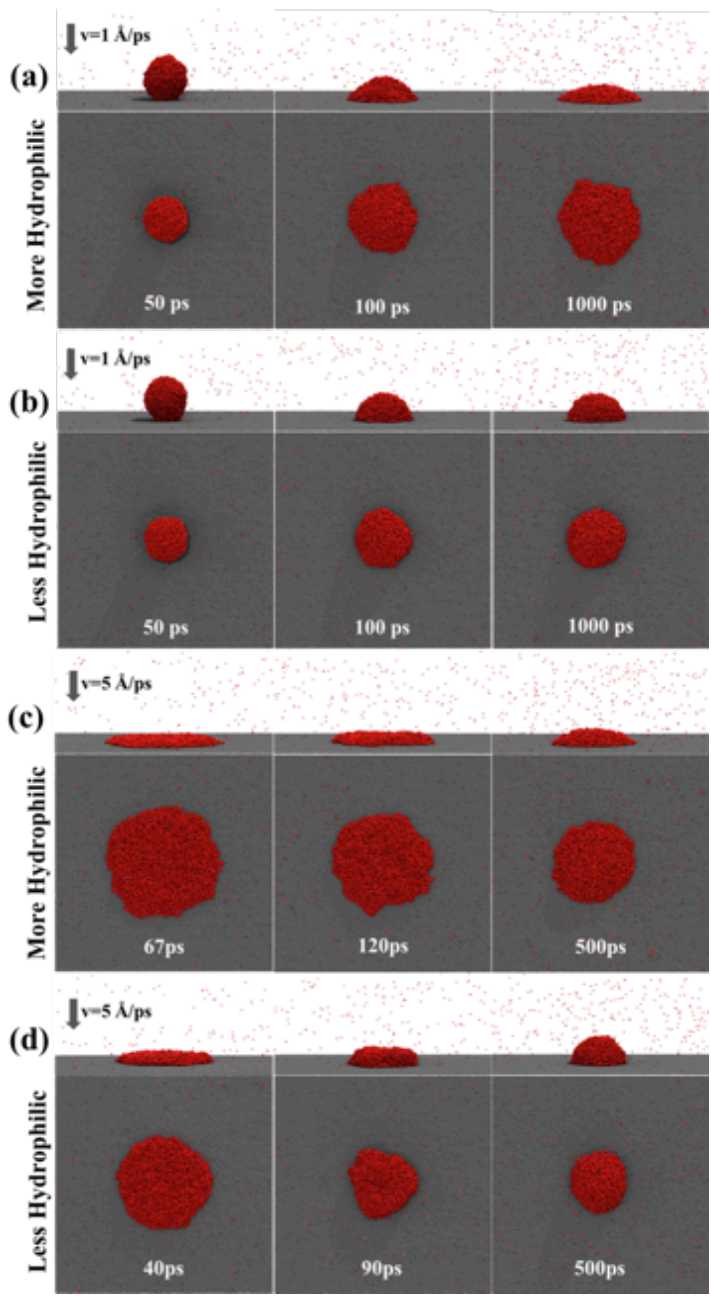

**Figure 2.** Snapshots from MD trajectories of the droplet impacting on graphene surface with different wettabilities. Impact at velocity of 1 Å/ps on; (a) more, and; (b) less hydrophilic surfaces. Impact at velocity of 5 Å/ps on; (c) more, and; (d) less hydrophilic surfaces.

For velocities larger than 1 Å/ps, as in Figure 2(c,d) for 5 Å/ps (see also Video 1 of Supporting Information), we noticed an intermediary state where the droplet first spreads up to a maximum diameter on the surface and then retracts to a smaller diameter size. We define this droplet diameter as the average of the droplet maximum lengths along x and y coordinates ($D_x$ and $D_y$ in Figure 3(a), respectively). The diameter at maximum spreading increases with the velocity of impact, as can be seen in Figures 3(b).

Graphene oscillates when the droplet impacts on its surface, showing higher amplitudes at moment of impact and vanishing as the simulations continues (Figure S1 on Supporting Information). These oscillations do not affect significantly the spreading patterns observed, as can be seen in a series of impact simulations with larger graphene sheets (Figures S2 on Supporting Information). It is also interesting to notice that the oscillation after impact dampers faster for the less hydrophilic surface. This happens because the less hydrophilic surface has less contact area with graphene, and the graphene oscillations affects the droplet energy less than the more hydrophilic surface, where more water molecules are into closer contact with graphene.

The diameters are always larger for the more hydrophilic surface than the less hydrophilic surface. The value of 11 Å/ps is the maximum impact velocity before the droplet fragment in the more hydrophilic surface, being 10 Å/ps for the less hydrophilic one. This difference in spreading is due to the difference in wettability between the two surfaces. After the maximum spread, the droplets retract, reducing their diameter contact with the surface. It is interesting to notice that for large simulation times (equilibrated configurations after impacts), the droplet spreading diameters are almost velocity independent. This is a consequence that the number of evaporated water molecules is not large enough to significantly affect the diameter values. The final droplet configuration is determined by the wettability set for graphene. The final diameter for the droplets on the more hydrophilic surface stays around 220 Å, with a contact angle of about 40º (Figure 3(c)). For the less hydrophilic surface, the final droplet diameters are around 140 Å, with contact angles of about 86º (Figure 3(d)).

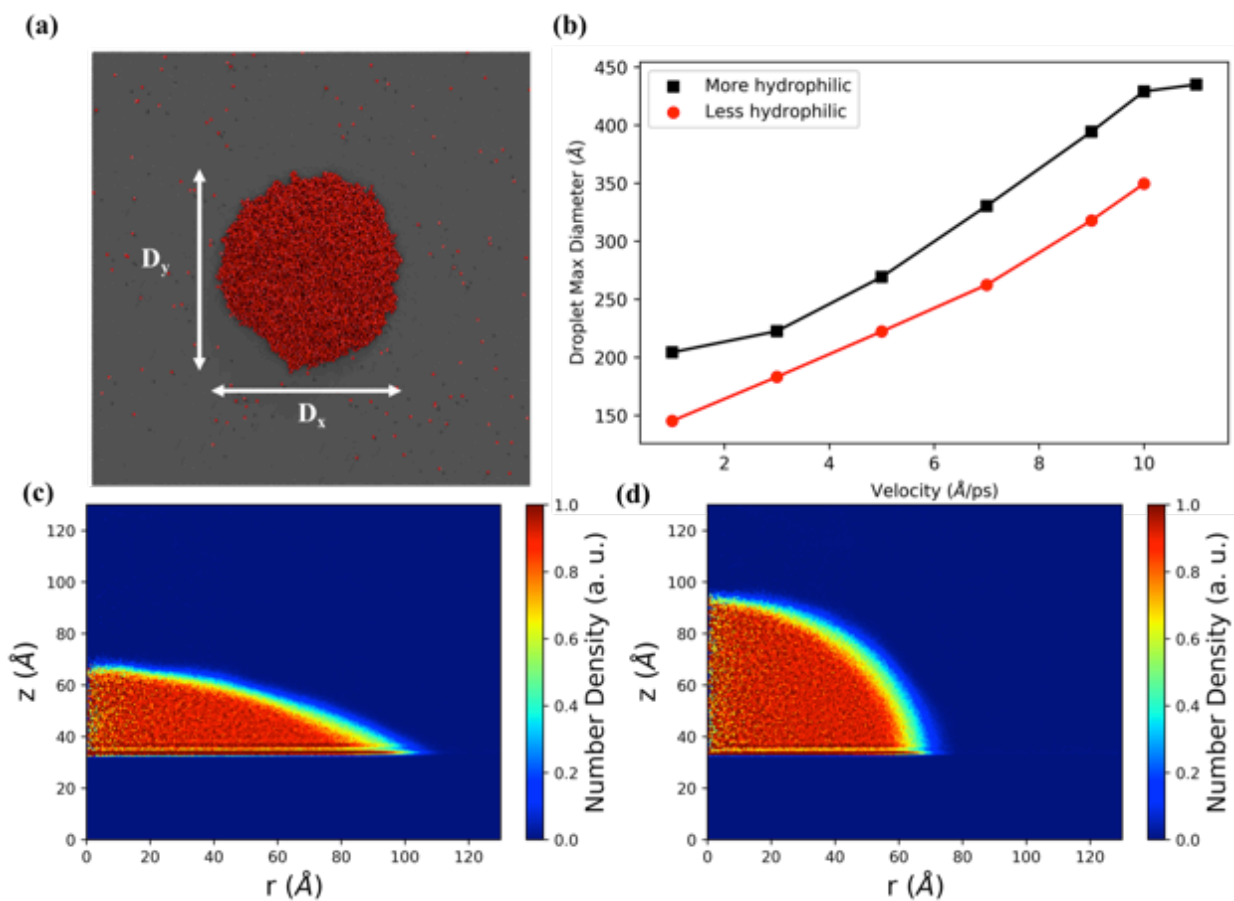

**Figure 3.** (a) Schematics of how the droplet diameter components along x and y directions were defined. (b) Droplet maximum diameter during spreading at different velocities. Density maps of droplet at equilibrium (after the impact at 10 Å/ps) on; (c) more, and; (d) less hydrophilic graphene surfaces.

*Energy Properties*

To better understand the patterns of spreading, we plot the potential energy of the droplets for the impact velocities of 1 and 5 Å/ps in Figure 4(a, b). For 1 Å/ps, the energy steadily increases as the droplet is being adhered on the surface, reaching a relatively stable energy after a few hundreds of picoseconds. The change in energy is due to the change from a spherical droplet prior to contact with graphene to a semispherical droplet on top of surface (Figure 2(a,b)). For 5

Å/ps impact velocity, during the spreading the potential energy presents a pronounced peak, that decreases as the droplet reaches a stable energy.

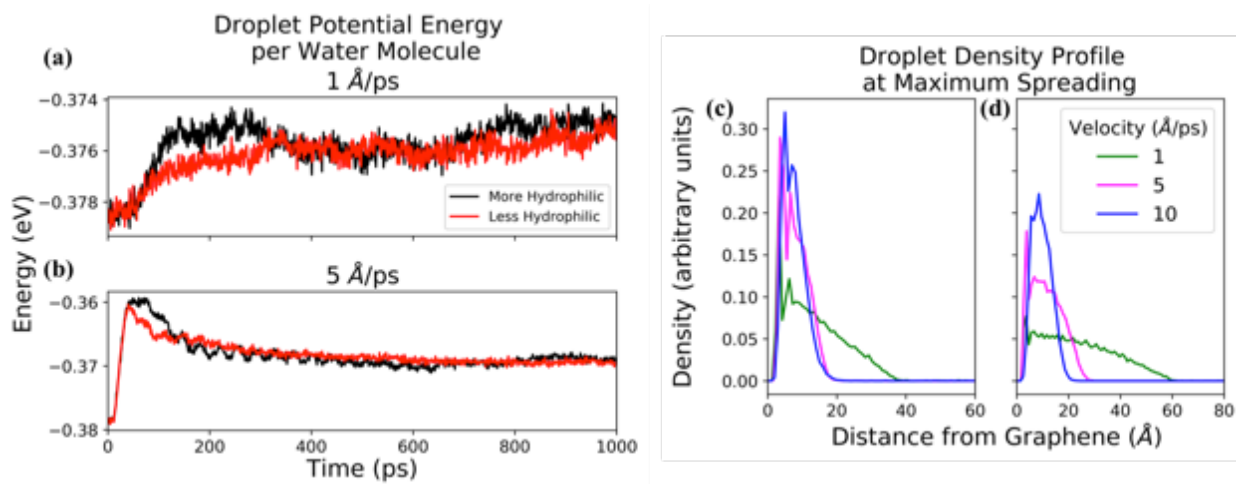

**Figure 4.** Temporal evolution of droplet potential energy per molecule for impact velocities of (a) 1 Å/ps and (b) 5 Å/ps. Droplet density profile along the perpendicular direction of graphene surface at selected impact velocities, for; (c) more, and; (d) less hydrophilic parametrizations.

This abrupt change in energy occurs because the droplet changes considerably its shape at maximum spreading when the velocity of impact increases. This can be observed in the density profiles at maximum spreading for the droplets in three different impact velocities for both parametrizations, as shown in Figure 4 (c, d). At 1 Å/ps, the droplet profiles show a well defined first solvation layer and the density disappears at 40 and 60 Å distance from graphene, respectively for the more and the less hydrophilic surface. At 5 Å/ps, the maximum spreading profiles still present the first solvation layer, but the droplets density increases nearby graphene surface. At 10 Å/ps, the whole droplet is condensed at less than 20 Å from the graphene surface, for both degrees of wettability.

The peak in potential energy (Figure 4(b)) that the droplet at 5 Å/ps reaches is similar for both parametrizations of graphene surface, but as the impact evolves the more hydrophilic surface maintains the droplet longer in an unstable formation than the less hydrophilic surface (Figure 2(c,d)). This occurs because that during spreading the droplet potential energy, due only to the interactions among water molecules, always increases at maximum spreading (Figure 5(a)). In the more hydrophilic surface, this instability is more pronounced, because the droplet can spread more and it is more affected by graphene surface. Besides that, during maximum spreading the graphene-droplet interaction energy is stronger (Figure 5(b)), as most of the molecules are closer to the surface. For a particular impact velocity, the more hydrophilic surface always presents stronger interaction with the droplet than the less hydrophilic one.

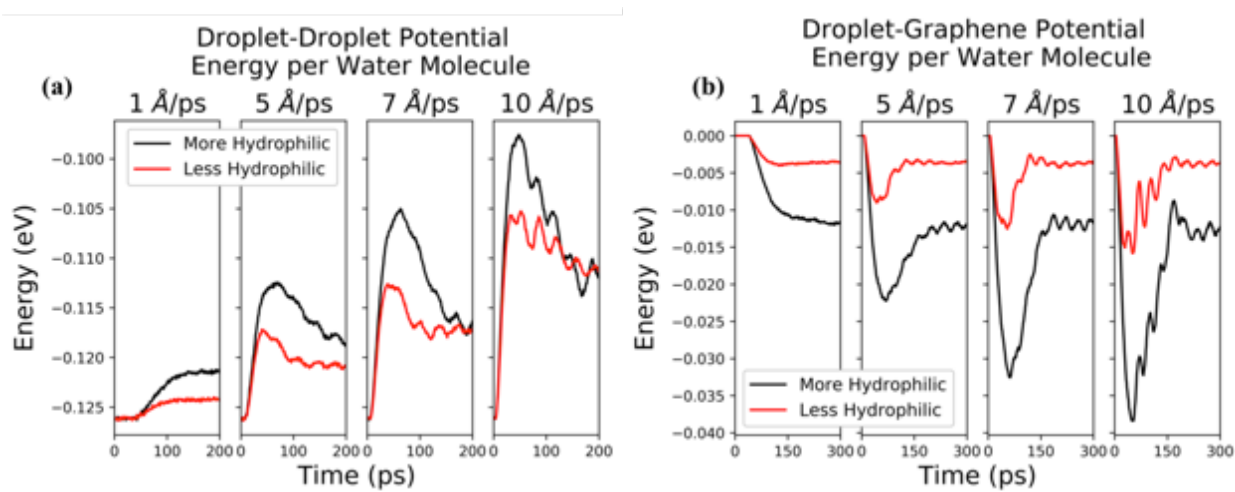

**Figure5.** (a) Droplet-droplet potential energy per water molecule for a set of impact velocities. (b) Droplet-graphene potential energy per water molecule for a set of impact velocities.

For 5 Å/ps and lower velocities, the droplet spreads homogeneously on the surface, with the contact area with graphene being similar to an expanding circle. This type of spreading makes the droplet at the less hydrophilic surface to rapidly recovers from the abrupt change in shape, reaching a more stable energy (and configuration) faster than the more hydrophilic surface, that interacts stronger with graphene. However, for impact velocity of 7 Å/ps, the spreading pattern changes to a more irregular one, and structural voids inside the structure of the droplet starts to appear (Figure 6 (a, b) and Video 2 of Supporting Information). This change in shape seems to be detrimental for the stability of the droplet at the less hydrophilic surface, as the voids make it harder for a fast recovery to a stable shape, resulting in a higher potential energy (more unstable). For even higher velocities, the deformation on the droplets increases for both parametrizations.

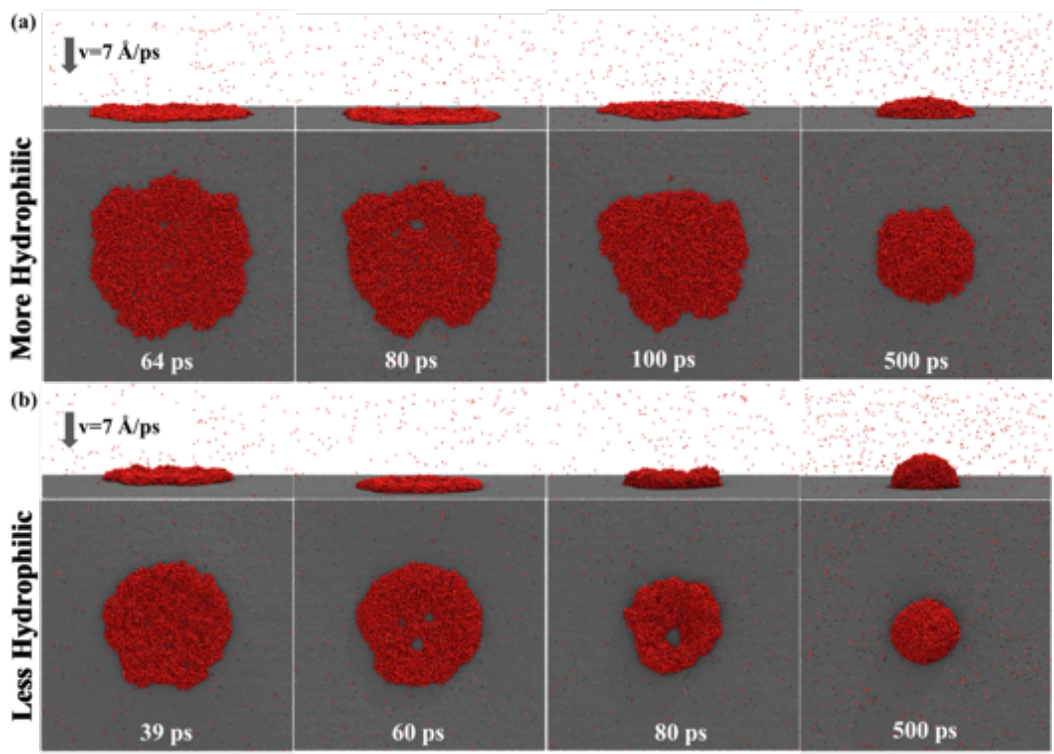

**Figure 6.** Snapshots from MD trajectories of the droplet impacting on graphene surface with different wettabilities. Impact at velocity of 7 Å/ps on; (a) more, and; (b) less hydrophilic surfaces.

Considering the case of the the potential energy of 7 Å/ps impact velocity (Figure 7(a)), at first for both parametrizations the peak in potential energy is the same. Just after impact the more hydrophilic surface maintains the droplet at that high-energy state longer than the less hydrophilic one. However, as the system evolves in time, the droplet energy at the more hydrophilic surface decreases faster than at the less hydrophilic surface. After some time, both configurations reach the same energy. As the velocity of impact increases, for example in Figure 7(b) for 10 Å/ps, this change in pattern becomes more pronounced resulting that the droplet at the less hydrophilic surface stays longer in an unstable configuration.

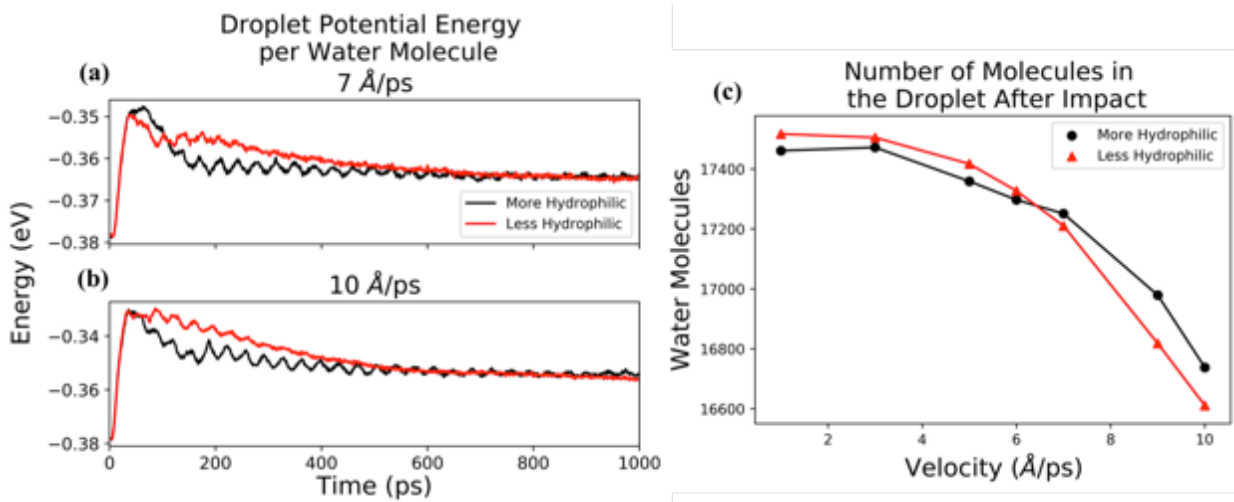

**Figure 7.** Droplet potential energy per water molecule at impact velocity of; (a) 7, and; (b) 10 Å/ps. (c) Number of molecules belonging to droplet after impact.

Despite surface wettability, the potential energy after impact always reaches about the same value for a determined velocity of impact, because in equilibrium the van der Waals and electrostatic interaction between water molecules counterbalances with the water-graphene interactions. Besides that, the amount of water molecules lost by the droplet during the impact is about the same for a determined velocity for both parametrizations (Figure 7(c)). The differences are considerably larger when we compare different velocities. Another important fact is that our simulations have constant number of atoms and volume, so eventually, vapor molecules nearby the droplet may return to it.

*Fragmentation of Droplets*

When the velocity of impact is increased to values beyond 11 Å/ps, for the more hydrophilic and 10 Å/ps for the less hydrophilic graphene, the droplet starts to fragment. As shown in Figure 8 (a), the droplet at the more hydrophilic surface has more resilience to fragmentation because it spreads more on the surface than the droplet at the less hydrophilic surface (Figure 8 (b) and Video 3 of Supporting Information).

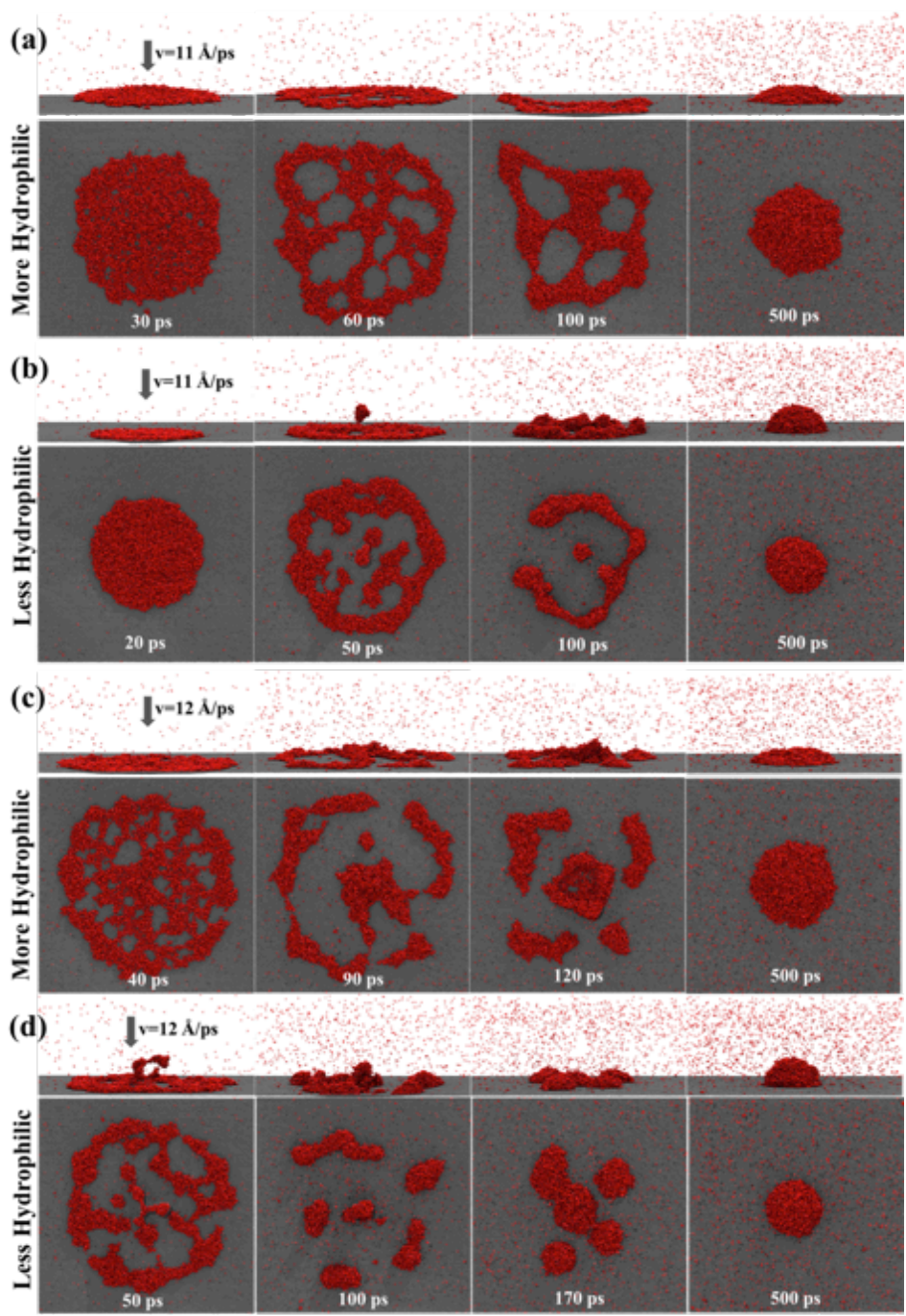

**Figure 8.** Snapshots from MD trajectories of droplets at 11 Å/ps on the (a) more and (b) less hydrophilic graphene surface. Snapshots from MD trajectories of droplets at 12 Å/ps on the (a) more and (b) less hydrophilic graphene surface.

The fragmentation starts with the formation of a large number of structural voids inside the droplet. As these voids continues to increase, eventually a small cluster of molecules becomes detached from the droplet. For the droplet at the more hydrophilic surface, the detachment occurs at the outer region of the droplet (Figure 8(c)). In the less hydrophilic surface, the detachment occurs first in the middle of the droplet, as can be seen in Figure 8 (b,d). This happens because the less hydrophilic surface has less interaction with the droplet, what makes the oscillations of graphene partially eject the liquid off the surface. The velocity of 11 Å/ps is the moment where this behavior starts to occur and becomes more pronounced as the velocity increases (Figure 8(d)). For the more hydrophilic surface this liquid ejection does not occur because the interaction graphene-droplet is stronger.

For velocities slightly larger than the limit of no-fragmentation, as the ones shown in Figure 8 (b,c,d), the small clusters of liquid detached during spreading are not ejected with high enough velocity, staying closer to the main droplet and eventually being rebounded to it. The fragmentation of the droplet occurs due to decreased thickness of liquid density during spreading at higher impact velocities. As can be seen in Figure 9 (a, b), at velocity of 3 Å/ps the density map of the droplet at maximum spreading is well distributed in its inner part, with low densities only in the vapor-liquid interface. The lateral kinetic energy that the droplets acquire after impact with graphene increases as the velocity of impact increases, for both parametrizations, as can be seen in Figure 8(c). Only after a few picoseconds (10 to 50 ps) after this maximum lateral kinetic energy is reached is that the maximum interaction energy between graphene and droplet occurs. This increase in lateral velocity makes the water layer during spreading thinner and wider as more downward velocity is applied (Figure 9 (d,e)), what results in a stronger interaction energy between liquid and surface (Figure 9 (f)).

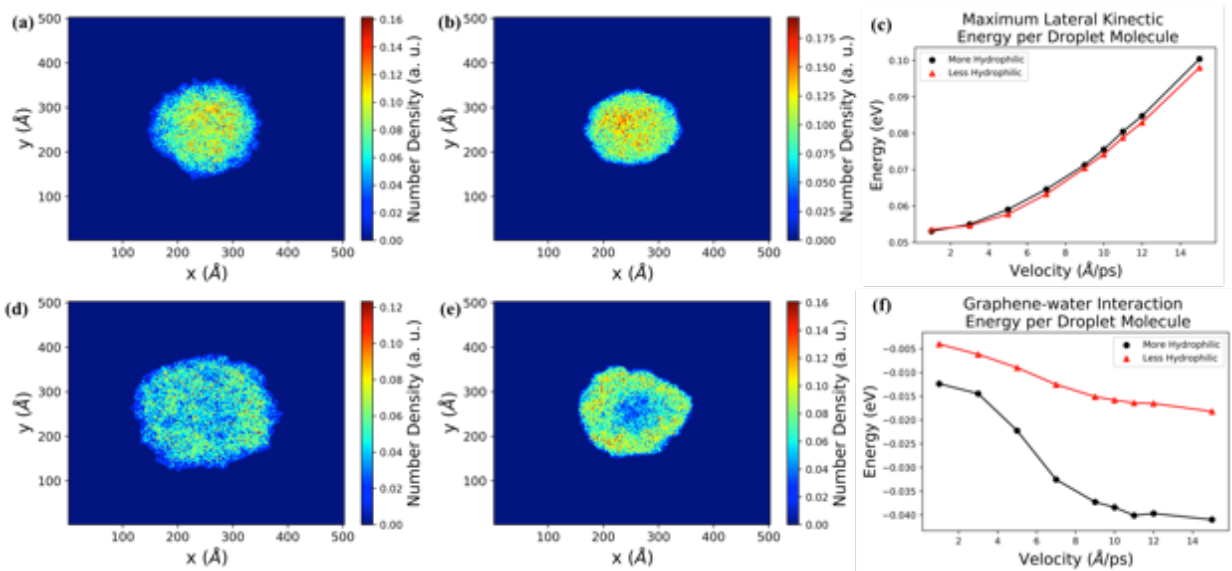

**Figure 9.** Density maps for droplets at 3 Å/ps for; (a) more, and; (b) less hydrophilic graphene. (c) Maximum lateral kinetic energy per droplet molecule. Density maps for droplets at 5 Å/ps for; (d) more, and; (e) less hydrophilic graphene. (f) Graphene-water interaction energy per droplet molecule at maximum spreading.

Even though this increase in lateral kinetic energy results that more water molecules move away from the center and move to the outer regions of the droplet, the interaction energy reaches a relative plateau after 7 Å/ps. This happens because at this point the thinner water layer starts to break out, with structural voids being formed and water concentrating more at the outer region (Figure 10 (a, b)). For higher impact velocities these voids become larger at maximum spread, with molecules accumulating into clusters (Figure 10 (c, d)). When the intermolecular interactions among these clusters are broken, the droplet starts to fragment.

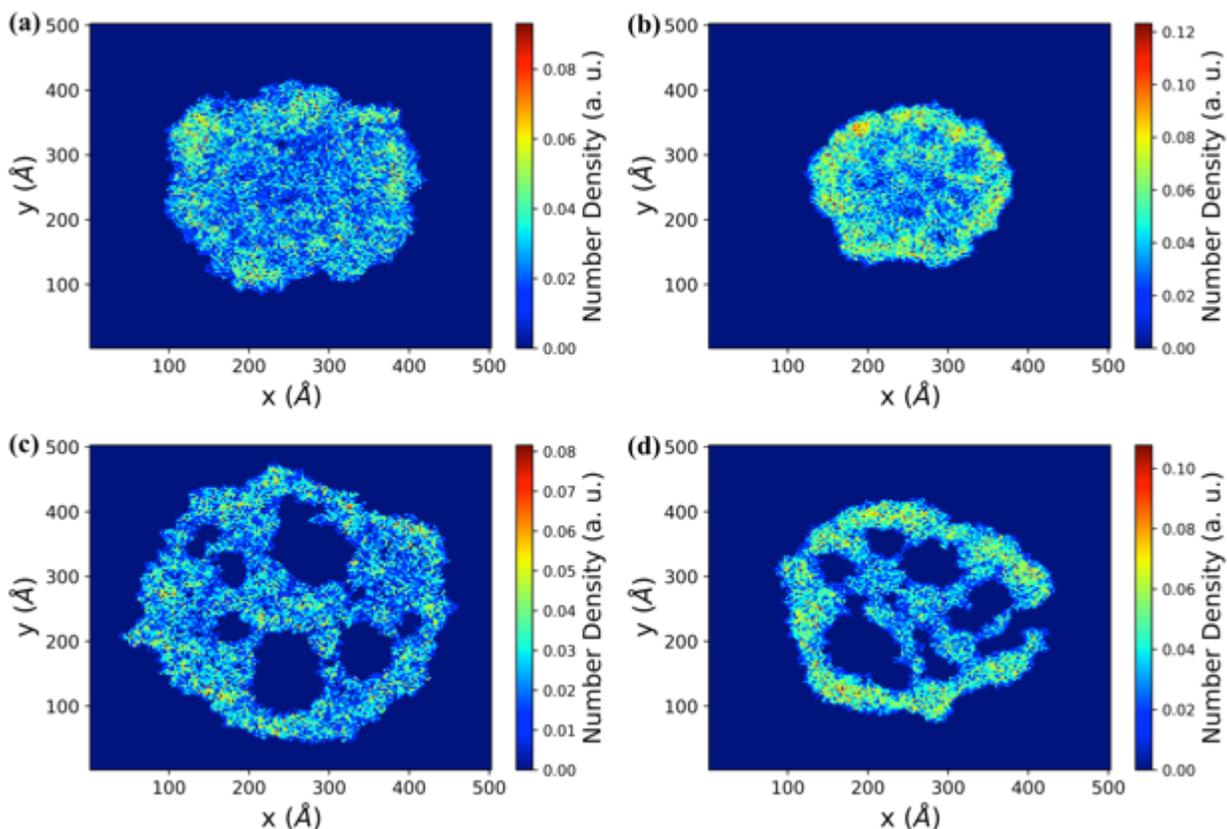

**Figure 10**. Density maps for droplets at 7 Å/ps for the (a) more and (b) less hydrophilic graphene. Density maps for droplets at 10 Å/ps for the (c) more and (d) less hydrophilic graphene.

**Conclusions**

This work demonstrated that water nanodroplets interacting with graphene at high impacting velocities, ranging from 1 to 15 Å/ps, have different spreading patterns depending on impact velocity and graphene wettability. Three ranges of velocities showed distinct patterns. For velocities up to 1 Å/ps the droplets are simply deposited on the surface, with the droplets reaching contact angles of 40º and 86º for the more and less hydrophilic graphene surfaces, respectively.

From 2 to 11 Å/ps droplets spread reaching a maximum diameter and rebound to an equilibrium configuration. The maximum diameter increases as the velocity of impact increases, but for a determined velocity this diameter is always large for the more hydrophilic surface. After a time of hundreds of picoseconds, the droplets equilibrate on the graphene surfaces, with their final contact angles and diameter having similar values as the ones of the simple deposition case (velocity smaller than 1 Å/ps).

During spreading, droplets present higher potential (more unstable) energies that decreases as they reach equilibrated configurations. Up to impact velocity of 6 Å/ps, the droplets spreading on the more hydrophilic surface stay longer at higher energy configuration. After this point, the spreading leads to structural droplet voids. These voids are more detrimental on the droplets interacting with the less hydrophilic graphene surface, which makes a significant change in how the curves of potential energy evolve. Because of the voids, the droplet on less hydrophilic surface stays longer on the unstable state than the more hydrophilic ones.

Increasing even more the velocity, for values beyond 11 (more hydrophilic case) and 10 (less hydrophilic case) the droplet fragments. The interaction energy between graphene-droplet shows that at these values the energy reaches a relative plateau, representing the maximum spreading that the droplet can reach. This spreading is a result of increasingly thinner layers of liquids, that after some points breaks into too many fragments.

**AUTHOR INFORMATION**

**Corresponding Author**

*galvao@ifi.unicamp.br

**Author Contributions**

The manuscript was written through contributions of all authors. All authors have given approval to the final version of the manuscript. ‡These authors contributed equally. (match statement to author names with a symbol).

**ACKNOWLEDGMENTS**

YMJ thanks São Paulo Research Foundation (FAPESP) Grant No. 2013/24500-2 for financial support. The authors acknowledge support from the Center for Computational Engineering and Sciences at University of Campinas (FAPESP/CEPID grant No. 2013/08293-7) and the Brazilian Agencies CNPq and CAPES.